\documentclass[11pt,preprint,showpacs,preprintnumbers,amsmath,amssymb]{revtex4}

\usepackage{exscale}
\usepackage{relsize}
\usepackage{graphicx}
\usepackage{dcolumn}
\usepackage{bm}
\usepackage{multirow}
\usepackage{CJK}
\usepackage{diagbox}
\usepackage{subfigure}

\begin{document}

\begin{CJK*}{GBK}{}

\title{Study of localized $CP$ violation in $B^-\rightarrow \pi^- \pi^+\pi^-$ and the branching ratio of $B^-\rightarrow \sigma(600)\pi^-$ in the QCD factorization approach}

\author{Jing-Juan Qi \footnote{e-mail: jjqi@mail.bnu.edu.cn}}
\affiliation{\small{Junior College, Zhejiang Wanli University, Zhejiang 315101, China}}

\author{Zhen-Yang Wang \footnote{e-mail: wangzhenyang@nbu.edu.cn}}
\affiliation{\small{Physics Department, Ningbo University, Zhejiang 315211, China}}

\author{Xin-Heng Guo \footnote{Corresponding author, e-mail: xhguo@bnu.edu.cn}}
\affiliation{\small{College of Nuclear Science and Technology, Beijing Normal University, Beijing 100875, China}}

\author{Zhen-Hua Zhang \footnote{Corresponding author, e-mail: zhangzh@usc.edu.cn}}
\affiliation{\small{School of Nuclear and Technology, University of South China, Hengyang, Hunan 421001, China}}

\author{Xian-Wei Kang \footnote{e-mail: 11112018023@bnu.edu.cn}}
\affiliation{\small{College of Nuclear Science and Technology, Beijing Normal University, Beijing 100875, China}}

\date{\today\\}

\begin{abstract}
In this work, within the QCD factorization approach, we study the localized integrated $CP$ violation in the $B^-\rightarrow \pi^-\pi^+\pi^-$ decay and the branching fraction of the $B^-\rightarrow\sigma\pi^-$ decay. Both the resonance and nonresonance contributions are included when we study the localized $CP$ asymmetry in the $B^-\rightarrow \pi^-\pi^+\pi^-$ decay. The resonance contributions from the scalar $\sigma(600)$ and vector $\rho^0(770)$ mesons are included. For the $\sigma(600)$ meson, we apply both the Breit-Wigner and Bugg models to deal with its propagator, and obtain $\mathcal{B}(B^-\rightarrow \sigma(600)\pi^-)<1.67\times10^{-6}$ and $\mathcal{B}(B^-\rightarrow \sigma(600) \pi^-)<1.946\times10^{-5}$ in these two models, respectively. We find that there is no allowed divergence parameters $\rho_S$ and $\phi_S$ to satisfy the experimental data $\mathcal{A_{CP}}(\pi^-\pi^+\pi^-)=0.584\pm0.082\pm0.027\pm0.007$ in the region $m_{\pi^+\pi^- \mathrm{high}}^2>15$ $\mathrm{GeV}^2$ and $m_{\pi^+\pi^-\mathrm{low}}^2<0.4$ $\mathrm{GeV}^2$ and the upper limit of $\mathcal{B}(B^-\rightarrow \sigma(600)\pi^-)$ in the Breit-Wigner model, however, there exists the region $\rho_S\in[1.70,3.34]$ and $\phi_S \in [0.50,4.50]$ satisfying the data for $\mathcal{A_{CP}}(\pi^-\pi^+\pi^-)$ and the upper limit of $\mathcal{B}(B^-\rightarrow \sigma(600)\pi^-)$ in the Bugg model. This reveals that the Bugg model is more plausible than the Breit-Wigner model to describe the propagator of the $\sigma(600)$ meson even though the finite width effects are considered in both models. The large values of $\rho_S$ indicate that the contributions from weak annihilation and hard spectator scattering processes are both large, especially, the weak annihilation contribution should not be negleted for $B$ decays to final states including a scalar meson.
\end{abstract}

\pacs{11.30.Er,13.25.Hw,14.40.-n}
\maketitle
\end{CJK*}

\section{Introduction}
Charge-Parity ($CP$) violation is essential to our understanding of both particle physics and the evolution of the early universe. It is one of the most fundamental and important properties of weak interaction, and has gained extensive attentions ever since its first discovery in 1964 \cite{Christenson:1964fg}. In the Standard Model (SM), $CP$ violation is related to the weak complex phase in the Cabibbo-Kobayashi-Maskawa (CKM) matrix, which describes the mixing of different generations of quarks \cite{Cabibbo:1963yz, Kobayashi:1973fv}. Besides the weak phase, a large strong phase is also needed for a large \emph{CP} asymmetry. Generally, this strong phase is provided by QCD loop corrections and some phenomenological models.

In recent years, the charmless three-body decays of $B$ mesons have attracted much more attention, because by studying them the CKM parameters can be determined and the possible new physics effects beyond the SM can be search for. However, the three-body decays of $B$ mesons are more complicated than the two-body cases, because both resonance and nonresonance contributions are involved in the hadronic matrix elements. For the three-body $B$ decay $B\rightarrow M_1M_2M_3$, under the factorization hypothesis, one of the nonresonance contributions arises from the transitions $B\rightarrow M_1M_2$. The nonresonance background in the charmless three-body $B$ decays due to the transition $B\rightarrow M_1M_2$ has been studied extensively based on the heavy meson chiral perturbation theory (HMChPT). In order to apply this approach, both of the final-state pseudoscalars in the $B\rightarrow M_1M_2$ transition have to be soft \cite{Cheng:2013dua}. Besides the nonresonance background, the three-body meson decays are generally dominated by intermediate resonances, namely, they proceed via quasi-two-body decays containing resonance states. LHCb has observed the large $CP$ asymmetry in the $B^-\rightarrow \pi^-\pi^+\pi^-$ decay in the localized region of the phase space \cite{Aaij:2013bla}, $\mathcal{A_{CP}}(\pi^-\pi^+\pi^-)=0.584\pm0.082\pm0.027\pm0.007$, for $m_{\pi^+\pi^- \mathrm{high}}^2>15$ $\mathrm{GeV}^2$ and $m_{\pi^+\pi^-\mathrm{low}}^2<0.4$ $\mathrm{GeV}^2$, which spans the $\sigma(600)$ and $\rho^0(770)$ mesons. In 2005, $BABAR$ Collaboration reported the upper limit of $\mathcal{B}(B^-\rightarrow \sigma \pi^-, \sigma\rightarrow \pi^+\pi^-)$ as $4.1\times10^{-6}$ \cite{Aubert:2005sk}. Both of these experimental results are important for us to study the $B$ decays including the scalar meson $\sigma(600)$.

Theoretically, to calculate the hadronic matrix elements of $B$ nonleptonic weak decays, some approaches, including the naive factorization \cite{Wirbel:1985ji,Bauer:1986bm}, the QCD factorization (QCDF) \cite{Beneke:2003zv,Beneke:2001ev}, the perturbative QCD (PQCD) approach \cite{Keum:2000ph}, and the soft-collinear effective theory (SCET) \cite{Bauer:2000ew}, have been developed and extensively employed in recent years. In this work, within the framework of QCDF, we will study the decays of $B^-\rightarrow \pi^-\pi^+\pi^-$ and $B^-\rightarrow \sigma\pi^-$.

The remainder of this paper is organized as follows. In Sect. ${\mathrm{\uppercase\expandafter{\romannumeral2}}}$, we present the form factors, decay constants and distribution amplitudes of different mesons. In Sect. ${\mathrm{\uppercase\expandafter{\romannumeral3}}}$, we present the formalism for $B$ decays in the QCDF approach. In Sect. ${\mathrm{\uppercase\expandafter{\romannumeral4}}}$, we give the calculations of the localized $CP$ violation and the branching ratio of the $B$ meson decays. The numerical results are given in Sect. ${\mathrm{\uppercase\expandafter{\romannumeral5}}}$ and we summarize our work in Sect ${\mathrm{\uppercase\expandafter{\romannumeral6}}}$.

\section{FORM FACTORS, DECAY CONSTANTS AND LIGHT-CONE DISTRIBUTION AMPLITUDES}
Since the form factors for $B\rightarrow P$, $B\rightarrow V$ and $B\rightarrow S$ ($P$, $V$ and $S$ represent pseudoscalar, vector and scalar mesons, respectively) weak transitions and light-cone distribution amplitudes and decay constants of $P$, $V$ and $S$ will be used in treating $B$ decays, we first discuss them in this section.

The form factors of $B$ to a meson weak transition can be decomposed as \cite{Wirbel:1985ji,Cheng:2010yd}
\begin{equation}\label{bv}
\begin{split}
\langle P(p')|\hat{V}_\mu|B(p)\rangle&=\bigg(p_\mu-\frac{m_B^2-m_P^2}{q^2}q_\mu\bigg) F_1^{BP}(q^2)+\frac{m_B^2-m_P^2}{q^2}q_\mu F_0^{BP}(q^2),\\
\langle V(p')|\hat{V}_\mu|B(p)\rangle&=\frac{2}{m_B+m_V}\varepsilon_{\mu\nu\rho\sigma} \epsilon^{*\nu} p^\rho p'^\sigma V^{BV}(q^2),\\
\langle V(p')|\hat{A}_\mu|B(p)\rangle&=i\bigg\{(m_B+m_V)\epsilon_\mu^*A_1^{BV}(q^2)-\frac{\epsilon^*\cdot q}{m_B+m_V}P_\mu  A_2^{BV}(q^2)\\
&-2m_V\frac{\epsilon^*\cdot P}{q^2} q_\mu [A_3^{BV}(q^2)- A_0^{BV}(q^2)]\bigg\},\\
\langle S(p')|\hat{A}_\mu|B(p)\rangle&=-i\bigg[\bigg(P_\mu-\frac{m_B^2-m_S^2}{q^2}q_\mu\bigg)F_1^{BS}(q^2)+\frac{m_B^2-m_S^2}{q^2}q_\mu F_0^{BS}(q^2)\bigg],\\
 \end{split}
\end{equation}
where $P_\mu=(p+p')_\mu$, $q_\mu=(p-p')_\mu$, $\hat{V}_\mu$, $\hat{A}_\mu$ and $\hat{S}_\mu$ are the weak vector, axial-vector and scalar currents, respectively, i.e. $\hat{V}_\mu=\bar{q}_f\gamma_\mu b, \hat{A}_\mu=\bar{q}_f\gamma_\mu \gamma_5b, \hat{S}=\bar{q}_f b$ with $q_f$ being the quark generated from the $b$ quark decay$, \epsilon_\mu$ is the polarization vector of $V$, $F_i^{BP}(q^2)$ $(i=0,1)$ and $A_i^{BV}(q^2)$ $(i=0,1,2,3)$ are the weak form factors. The form factors included in Eq. (\ref{bv}) satisfy $F_1^{BP}(0)=F_0^{BP}(0)$, $A_3^{BV}(0)=A_0^{BV}(0)$, $A_3^{BV}(q^2)=[(m_B+m_{V})/(2m_{V})]A_1^{BV}(q^2)-[(m_B+m_{V})/(2m_{V})]A_2^{BV}(q^2)$ and $F_1^{BS}(q^2)=F_0^{BS}(q^2)$.

 The decay constants are defined as \cite{Cheng:2010yd}
\begin{equation}\label{dc}
\begin{split}
\langle P(p')|\hat{A}_\mu|0\rangle&=-if_P p'_\mu,\\
\langle V(p')|\hat{V}_\mu|0\rangle&=f_Vm_V\epsilon^*_\mu, \quad\langle V(p')|\overline{q}\sigma_{\mu\nu}q'|0\rangle=f_V^\perp (p'_\mu\epsilon^*_\nu-p'_\nu\epsilon^*_\mu)m_V,\\
\langle S(p')|\hat{V}_\mu|0\rangle&=f_S p'_\mu, \quad \langle S(p')|\hat{S}|0\rangle=m_S\bar{f}_S.\\
\end{split}
\end{equation}

The twist-2 light-cone distribution amplitudes (LCDA) for the pseudoscalar and vector mesons are respectively \cite{Beneke:2003zv,Cheng:2010yd}
\begin{equation}\label{phiM}
\Phi_{M}(x,\mu)=6x(1-x)\bigg[\sum\limits_{m=0}^\infty \alpha_m^{M}(\mu)C_m^{3/2}(2x-1)\bigg], \quad M=P,V,
 \end{equation}
and the twist-3 ones are respectively
 \begin{equation}\label{phim}
\begin{split}
\Phi_m(x)=
\begin{cases}
1,& \quad m=p,  \\
3\bigg[2x-1+\sum\limits_{m=1}^\infty \alpha_{m,\perp}^V(\mu)P_{m+1}(2x-1)\bigg],& \quad m=v, \\
\end{cases}
\end{split}
 \end{equation}
where $C_m^{3/2}$ and $P_m$ are the Gegenbauer and Legendre polynomials in Eq. (\ref{phiM}) and Eq. (\ref{phim}), respectively, $\alpha^M_m(\mu)$ and $\alpha_{m,\perp}^V(\mu)$ are Gegenbauer moments which depend on the scale $\mu$.

 The twist-2 light-cone distribution amplitude for a scalar meson reads \cite{Cheng:2005nb,Cheng:2007st}
\begin{equation}\label{PhiS}
\Phi_S(x,\mu)^{(n,s)}=\bar{f}^{n,s}_S6x(x-1)\sum_{m=1,3,5}^\infty B_m(\mu)C_m^{3/2}(2x-1),
 \end{equation}
 where $B_m$ are Gegenbauer moments, $\bar{f}_S$ is the decay constant of the scalar meson, $n$ denotes the $u$, $d$ quark component of the scalar meson, $n=\frac{1}{\sqrt{2}}(u\bar{u}+d\bar{d})$, and $s$ denotes the component $s\bar{s}$. As for the twist-3 ones, we shall take the asymptotic forms \cite{Cheng:2005nb,Cheng:2007st}
 \begin{equation}
 \Phi_s(x)^{(n,s)}=\bar{f}^{n,s}_S.
 \end{equation}
\section{B DECAYS IN QCD FACTORIZATION}
In the SM, the effective weak Hamiltonian for non-leptonic $B$-meson decays is given by \cite{Buchalla:1995vs}
 \begin{equation}\label{Hamiltonian}
 \mathcal{H}_{eff}=\frac{G_F}{\sqrt{2}}\bigg[\sum_{p=u,c}\sum_{D=d,s}\lambda_{p}^{(D)}(c_1O_1^p+c_2Q_2^p+\sum_{i=3}^{10}c_iO_i+c_{7\gamma}O_{7\gamma}+c_{8g}O_{8g})\bigg]+h.c.,
 \end{equation}
 where $\lambda_p^{(D)}=V_{pb}V_{pD}^*$, $V_{pb}$ and $V_{pD}$ are the CKM matrix elements, $G_F$ represents the Fermi constant, $c_i$ $(i=1-10,7\gamma,8g)$ are Wilson coefficients, $O_{1,2}^p$ are the tree level operators, $O_{3-6}$ are the QCD penguin operators, $O_{7-10}$ arise from electroweak penguin diagrams, and $O_{7\gamma}$ and$O_{8g}$ are the electromagnetic and chromomagnetic dipole operators, respectively.

Within the framework of QCD factorization \cite{Beneke:2003zv,Beneke:2001ev}, the matrix elements of the effective Hamiltonian are written in the form
\begin{equation}
\langle{M_1M_2}|\mathcal{H}_{eff}|B\rangle=\sum_{p=u,c}\lambda_{p}^{(D)}\langle{M_1M_2}|\mathcal{T}_A^p+\mathcal{T}_B^p|B\rangle,
\end{equation}
where $\mathcal{T}_A^p$ describes the contribution from naive factorization, vertex correction, penguin amplitude and spectator scattering expressed in terms of the parameters $a_i^p$, while $\mathcal{T}_B^p$ contains annihilation topology amplitudes characterized by the annihilation parameters $b_i^p$.

The flavor parameters $a_i^p$ are basically the Wilson coefficients in conjunction with short-distance nonfactorizable corrections such as vertex corrections and hard spectator interactions. In general, they have the following expressions \cite{Beneke:2003zv}:
\begin{equation}\label{a}
\begin{split}
a_i^p{(M_1M_2)}&={(c'_i+\frac{c'_{i\pm1}}{N_c})}N_i{(M_2)}+\frac{c'_{i\pm1}}{N_c}\frac{C_F\alpha_s}{4\pi}{\bigg[V_i{(M_2)}+\frac{4\pi^2}{N_c}H_i{(M_1M_2)}\bigg]+P_i^p{(M_2)}},
\end{split}
 \end{equation}
where $c'_i$ are effective Wilson coefficients which are defined as $c_i(m_b)\langle O_i(m_b)\rangle=c'_i\langle O_i\rangle^{\mathrm{tree}}$, with $\langle O_i\rangle^{\mathrm{tree}}$ being the matrix element at the tree level, the upper (lower) signs apply when $i$ is odd (even), $N_i{(M_2)}$ is leading-order coefficient, $C_F={(N_c^2-1)}/{2N_c}$ with $N_c=3$, the quantities $V_i{(M_2)}$ account for one-loop vertex corrections, $H_i{(M_1M_2)}$ describe hard spectator interactions with a hard gluon exchange between the emitted meson and the spectator quark of the $B$ meson, and $P_i^p{(M_1M_2)}$ are from penguin contractions \cite{Beneke:2003zv}.

The expressions of the quantities $N_i(M_2)$ read
\begin{equation}
\begin{split}
N_i{(V)}=
\begin{cases}
0& \quad i=6,8,  \\
1& \quad \text{else}, \\
\end{cases}\quad N_i{(S)}=1,\quad N_i{(P)}=1.
\end{split}
 \end{equation}

 When $M_1M_2=VP,PV$, the correction from the hard gluon exchange between $M_2$ and the spectator quark is given by \cite{Beneke:2003zv,Beneke:2001ev}
 \begin{equation}\label{H1}
\begin{split}
 H_i{(M_1M_2)}&=\frac{f_Bf_{M_1}}{2m_V\epsilon_{V}^*\cdot p_BF_0^{B\rightarrow M_1}(0)}\int_{0}^{1}\frac{d\xi}{\xi}\Phi_{B}{(\xi)}\int_{0}^{1}dx\int_{0}^{1}dy{\bigg[\frac{\Phi_{M_2}{(x)}\Phi_{M_1}{(y)}}{\bar{x}\bar{y}}+{r_\chi^{M_1}}\frac{\Phi_{M_2}{(x)}\Phi_{m_1}{(y)}}{x\bar{y}}\bigg]},
\end{split}
\end{equation}
for $i=1-4,9,10$,
\begin{equation}\label{H2}
\begin{split}
 H_i{(M_1M_2)}&=-\frac{f_Bf_{M_1}}{2m_V\epsilon_{V}^*\cdot p_BF_0^{B\rightarrow M_1}(0)}\int_{0}^{1}\frac{d\xi}{\xi}\Phi_{B}{(\xi)}\int_{0}^{1}dx\int_{0}^{1}dy{\bigg[\frac{\Phi_{M_2}{(x)}\Phi_{M_1}{(y)}}{{x}\bar{y}}+{r_\chi^{M_1}}\frac{\Phi_{M_2}{(x)}\Phi_{m_1}{(y)}}{\bar{x}\bar{y}}\bigg]},
\end{split}
\end{equation}
for $i=5,7$ and $H_i(M_1M_2)=0$ for $i=6,8$.

When $M_1M_2=SP,PS$ \cite{Beneke:2003zv,Cheng:2005nb,Cheng:2007st},
\begin{equation}\label{H3}
\begin{split}
 H_i{(M_1M_2)}&=\frac{f_B}{F_0^{B\rightarrow M_1}m_B^2}\int_{0}^{1}\frac{d\xi}{\xi}\Phi_{B}{(\xi)}\int_{0}^{1}dx\int_{0}^{1}dy{\bigg[\frac{\Phi_{M_2}{(x)}\Phi_{M_1}{(y)}}{\bar{x}\bar{y}}+{r_\chi^{M_1}}\frac{\Phi_{M_2}{(x)}\Phi_{m_1}{(y)}}{x\bar{y}}\bigg]},
\end{split}
\end{equation}
for $i=1-4,9,10$,
\begin{equation}\label{H4}
\begin{split}
 H_i{(M_1M_2)}&=-\frac{f_B}{F_0^{B\rightarrow M_1}m_B^2}\int_{0}^{1}\frac{d\xi}{\xi}\Phi_{B}{(\xi)}\int_{0}^{1}dx\int_{0}^{1}dy{\bigg[\frac{\Phi_{M_2}{(x)}\Phi_{M_1}{(y)}}{{x}\bar{y}}+{r_\chi^{M_1}}\frac{\Phi_{M_2}{(x)}\Phi_{m_1}{(y)}}{\bar{x}\bar{y}}\bigg]},
\end{split}
\end{equation}
for $i=5,7$ and $H_i(M_1M_2)=0$ for $i=6,8$.

In Eqs. (\ref{H1}-\ref{H4}) $\bar{x}=1-x$, $\bar{y}=1-y$, and $r_\chi^{M_i}$ $(i=1,2)$ are ``chirally-enhanced" terms which are defined as
\begin{equation}
\begin{split}
r_\chi^{P}(\mu)&=\frac{2m_P^2}{m_b(\mu)(m_{q_1}+m_{q_2})(\mu)},\quad r_\chi^{V}=\frac{2m_{V}}{m_b(\mu)}\frac{f_{V}^\bot(\mu)}{f_{V}},\\
r_\chi^{S}&=\frac{2m_S}{m_b(\mu)}\frac{\bar{f}_S(\mu)}{f_S}=\frac{2m_S^2}{m_b(\mu)(m_2(\mu)-m_1(\mu))},\quad \bar{r}_\chi^{S}=\frac{2m_S}{m_b(\mu)}.\\
\end{split}
\end{equation}

The weak annihilation contributions to $B\rightarrow M_1M_2$ can be described in terms of $b_i$ and $b_{i,EW}$, which have the following expressions:
\begin{equation}\label{b}
\begin{split}
&b_1=\frac{C_F}{N_c^2}c'_1A_1^i, \quad b_2=\frac{C_F}{N_c^2}c'_2A_1^i, \\
&b_3^p=\frac{C_F}{N_c^2}\bigg[c'_3A_1^i+c'_5(A_3^i+A_3^f)+N_cc'_6A_3^f \bigg],\quad b_4^p=\frac{C_F}{N_c^2}\bigg[c'_4A_1^i+c'_6A_2^i \bigg], \\
&b_{3,EW}^p=\frac{C_F}{N_c^2}\bigg[c'_9A_1^i+C'_7(A_3^i+A_3^f)+N_cc'_8A_3^f \bigg],\\
&b_{4,EW}^p=\frac{C_F}{N_c^2}\bigg[c'_{10}A_1^i+c'_8A_2^i \bigg],
\end{split}
 \end{equation}
where the subscripts 1, 2, 3 of $A_n^{i,f}(n=1,2,3)$ stand for the annihilation amplitudes induced from $(V-A)(V-A)$, $(V-A)(V+A)$, and $(S-P)(S+P)$ operators, respectively, the superscripts $i$ and $f$ refer to gluon emission from the initial- and final-state quarks, respectively. Their explicit expressions are given by \cite{Beneke:2003zv,Cheng:2010yd,Cheng:2010hn,Cheng:2005nb,Cheng:2007st}
\begin{equation}\label{Ai}
\begin{split}
 A_1^i&=\pi\alpha_s\int_0^1 dx dy\begin{cases}
 \bigg(\Phi_{M_2}(x)\Phi_{M_1}(y)\bigg[\frac{1}{y(1-x\bar{y})}+\frac{1}{\bar{x}^2y}\bigg]-r_\chi^{M_1}r_\chi^{M_2} \Phi_{m_2}(x)\Phi_{m_1}(y)\frac{2}{\bar{x}y}\bigg),\quad \text{for $M_1M_2=VP,PS,$}\\
 \bigg(\Phi_{M_2}(x)\Phi_{M_1}(y)\bigg[\frac{1}{y(1-x\bar{y})}+\frac{1}{\bar{x}^2y}\bigg]+r_\chi^{M_1}r_\chi^{M_2} \Phi_{m_2}(x)\Phi_{m_1}(y)\frac{2}{\bar{x}y}\bigg),\quad \text{for $M_1M_2=PV,SP,$}\\
 \end{cases}\\
 A_2^i&=\pi\alpha_s\int_0^1 dx dy\begin{cases}
 \bigg(-\Phi_{M_2}(x)\Phi_{M_1}(y)\bigg[\frac{1}{\bar{x}(1-x\bar{y})}+\frac{1}{\bar{x}y^2}\bigg]+r_\chi^{M_1}r_\chi^{M_2} \Phi_{m_2}(x)\Phi_{m_1}(y)\frac{2}{\bar{x}y}\bigg),\quad \text{for $M_1M_2=VP,PS,$}\\
 \bigg(-\Phi_{M_2}(x)\Phi_{M_1}(y)\bigg[\frac{1}{\bar{x}(1-x\bar{y})}+\frac{1}{\bar{x}y^2}\bigg]-r_\chi^{M_1}r_\chi^{M_2} \Phi_{m_2}(x)\Phi_{m_1}(y)\frac{2}{\bar{x}y}\bigg),\quad \text{for $M_1M_2=PV,SP,$}\\
 \end{cases}\\
 A_3^i&=\pi\alpha_s\int_0^1 dx dy\begin{cases}
 \bigg(r_\chi^{M_1}\Phi_{M_2}(x)\Phi_{m_1}(y)\frac{2\overline{y}}{\overline{x}y(1-x\bar{y})}+r_\chi^{M_2} \Phi_{M_1}(y)\Phi_{m_2}(x)\frac{2x}{\overline{x}y(1-x\bar{y})}\bigg),\quad \text{for $M_1M_2=VP,PS,$}\\
 \bigg(-r_\chi^{M_1}\Phi_{M_2}(x)\Phi_{m_1}(y)\frac{2\overline{y}}{\overline{x}y(1-x\bar{y})}+r_\chi^{M_2} \Phi_{M_1}(y)\Phi_{m_2}(x)\frac{2x}{\overline{x}y(1-x\bar{y})}\bigg),\quad \text{for $M_1M_2=PV,SP,$}\\
 \end{cases}\\
 A_3^f&=\pi\alpha_s\int_0^1 dx dy\begin{cases}
 \bigg(r_\chi^{M_1}\Phi_{M_2}(x)\Phi_{m_1}(y)\frac{2(1+\overline{x})}{\overline{x}^2y}-r_\chi^{M_2} \Phi_{M_1}(y)\Phi_{m_2}(x)\frac{2(1+y)}{\overline{x}y^2}\bigg),\quad \text{for $M_1M_2=VP,PS,$}\\
 \bigg(-r_\chi^{M_1}\Phi_{M_2}(x)\Phi_{m_1}(y)\frac{2(1+\overline{x})}{\overline{x}^2y}-r_\chi^{M_2} \Phi_{M_1}(y)\Phi_{m_2}(x)\frac{2(1+y)}{\overline{x}y^2}\bigg),\quad \text{for $M_1M_2=PV,SP,$}\\
 \end{cases}\\
 A_1^f&=A_2^f=0.\\
 \end{split}
  \end{equation}

When dealing with the weak annihilation contributions and the hard spectator contributions, one has to deal with the infrared endpoint singularity $X=\int_0^1 dx/(1-x)$. The treatment of this endpoint divergence is model dependent, and we follow Ref. \cite{Beneke:2003zv} to parameterize the endpoint divergence in the annihilation and hard spectator diagrams as
\begin{equation}\label{XH}
X_{A(H)}^{M_1M_2}=(1+\rho^{M_1M_2}_{A(H)} e^{i\phi_{A(H)}^{M_1M_2}})\ln\frac{m_B}{\Lambda_h},
\end{equation}
where $\Lambda_h$ is a typical scale of order 0.5 $\mathrm{GeV}$, $\rho_{A(H)}^{M_1M_2}$ is an unknown real parameter and $\phi_{A(H)}^{M_1M_2}$ is a free strong phase in the range $[0,2\pi]$ for the annihilation (hard spectator) process. In our work, we will adopt the $X_H^{M_1M_2}=X_A^{M_1M_2}=X^{M_1M_2}$ for the $B\rightarrow PV$ decays \cite{Wang:2016yrm,Cheng:2009cn,Cheng:2010yd}, but for the $B\rightarrow SP$ decays, we will follow our earlier work \cite{Qi:2018wkj,Qi:2018syl} to use a further assumption $X^{M_1M_2}=X^{M_2 M_1}$ compared with the $B\rightarrow PV$ decays.

\section{CALCULATION OF CP VIOLATION AND BRANCHING RATIO}

\subsection{Nonresonance background}

In the absence of resonances, the factorizable nonresonance amplitude for the $B^-\rightarrow \pi^-\pi^+\pi^-$ decay has the following expression \cite{Cheng:2007si,Cheng:2013dua}:
\begin{equation}\label{ANR}
A_{NR}=\frac{G_F}{\sqrt{2}}\sum_{p=u,c}\lambda_{p}^{d}\langle\pi^+(p_1)\pi^-(p_2)|(\bar{u}b)_{V-A}|B^-\rangle\langle \pi^-(p_3)|(\bar{d}u)_{V-A}|0\rangle[a_1\delta_{pu}+a_4^p+a_{10}^p-(a_6^p+a_8^p)r_\chi^\pi].
\end{equation}
For the parameters $a_i$ which contain effective Wilson coefficients, we take the following values \cite{Cheng:2013dua,Cheng:2007si}:
\begin{equation}
\begin{split}
a_1&=0.99\pm0.037i,\quad a_2=0.19-0.11i,\quad a_3=-0.002+0.004i,\quad a_5=0.0054-0.005i,\\
a_4^u&=-0.03-0.02i,\quad a_4^c=-0.04-0.008i,\quad a_6^u=-0.006-0.02i,\quad a_6^c=-0.006-0.006i,\\
a_7&=0.54\times10^{-4}i,\quad a_8^u=(4.5-0.5i)\times10^{-4},\quad a_8^c=(4.4-0.3i)\times10^{-4},\\
a_9&=-0.010-0.0002i,\quad a_{10}^u=(-58.3+86.1i)\times10^{-5},\quad a_{10}^c=(-60.3+88.8i)\times10^{-5}.
\end{split}
\end{equation}
For the current-induced process, the amplitude $\langle\pi^+\pi^-|(\bar{u}b)_{V-A}|B^-\rangle\langle \pi^-|(\bar{d}u)_{V-A}|0\rangle$ can be expressed in terms of three unknown form factors \cite{Lee:1992ih,Cheng:2013dua,Cheng:2007si}
\begin{equation}\label{ANR1}
\begin{split}
A_{\mathrm{current-ind}}^{\mathrm{HMChPT}}&\equiv\langle\pi^+(p_1)\pi^-(p_2)|(\bar{u}b)_{V-A}|B^-\rangle\langle \pi^-(p_3)|(\bar{d}u)_{V-A}|0\rangle\\
&=-\frac{f_\pi}{2}[2m_3^2r+(m_B^2-s_{12}-m_3^2)\omega_++(s_{23}-s_{13}-m_2^2+m_1^2)\omega_-],\\
\end{split}
\end{equation}
where $r$ and $\omega_{\pm}$ are form factors which have the expressions as \cite{Lee:1992ih,Fajfer:1998yc}
 \begin{equation}
\begin{split}
\omega_+&=-\frac{g}{f_\pi^2}\frac{f_{B^*}m_{B^*}\sqrt{m_Bm_{B^*}}}{s_{23}-m_{B^*}^2}\bigg[1-\frac{(p_B-p_1)\cdot p_1}{m_{B^*}^2}\bigg]+\frac{f_B}{2f_\pi^2},\\
\omega_-&=\frac{g}{f_\pi^2}\frac{f_{B^*}m_{B^*}\sqrt{m_Bm_{B^*}}}{s_{23}-m_{B^*}^2}\bigg[1+\frac{(p_B-p_1)\cdot p_1}{m_{B^*}^2}\bigg],\\
r&=\frac{f_B}{2f_\pi^2}-\frac{f_B}{f_\pi^2}\frac{p_B\cdot(p_2-p_1)}{(p_B-p_1-p_2)^2-m_{B^2}}\\
&+\frac{2gf_{B^*}}{f_\pi^2}\sqrt{\frac{m_B}{m_{B^*}}}\frac{(p_B-p_1)\cdot p_1}{s_{23}-m_{B^*}^2}-\frac{4g^2f_B}{f_\pi^2}\frac{m_Bm_{B^*}}{(p_B-p_1-p_2)^2-m_B^2}\\
&\times\frac{p_1\cdot p_2-p_1\cdot(p_B-p_1) p_2\cdot(p_B-p_1)/m_{B^*}^2}{s_{23}-m_{B^*}^2},
\end{split}
\end{equation}
where $s_{ij}\equiv(p_i+p_j)^2$, $g$ is a heavy-flavor-independent strong coupling which can be extracted from the CLEO measurement of the $D^{*+}$ decay width, $|g|=0.59\pm0.01\pm0.07$ \cite{Ahmed:2001xc}, which sign is fixed to be negative in Ref. \cite{Yan:1992gz}.

However, the predicted nonresonance results based on the HMChPT are not recovered in the soft meson region and lead to decay rates that are too large and in disagreement with experiment \cite{Cheng:2002qu}. This issue is related to the applicability of the HMChPT, which requires the two mesons in  the final state in the $B\rightarrow M_1M_2$ transition have to be soft and hence an exponential form of the amplitudes is necessary \cite{Cheng:2007si,Cheng:2016ajl},
\begin{equation}\label{ANR2}
A_{\mathrm{current-ind}}=A_{\mathrm{current-ind}}^{\mathrm{HMChPT}}e^{-\alpha_\mathrm{NR}p_B\cdot(p_1+p_2)}e^{i\mathrm{\phi_{12}}},
\end{equation}
where $\alpha_{\mathrm{NR}}=0.081_{-0.009}^{+0.015}\mathrm{GeV}^{-2}$, and the phase $\phi_{12}$ of the nonresonance amplitude will be set to zero for simplicity \cite{Cheng:2007si,Cheng:2016ajl}.

Combining Eqs. (\ref{ANR}) and (\ref{ANR1}-\ref{ANR2}), the decay amplitude via $B^-\rightarrow NR\rightarrow \pi^-\pi^+\pi^-$ can be finally written as
\begin{equation}\label{ANR3}
\begin{split}
A_{NR}&=\frac{G_F}{\sqrt{2}}\sum_{p=u,c}\lambda_{p}^{d}\bigg\{-\frac{f_\pi}{2}e^{-\alpha_\mathrm{NR}(m_B^2-m_\pi^2-s_{\pi\pi\mathrm{low}})/2}\bigg[a_1\delta_{pu}+a_4^p+a_{10}^p-(a_6^p+a_8^p)r_\chi^\pi\bigg]\\
&\times\bigg[2m_\pi^2r+\bigg(m_B^2-s_{\pi\pi\mathrm{low}}-m_\pi^2\bigg)\omega_++\bigg(m_B^2+3m_\pi^2-s_{\pi\pi\mathrm{low}}-2s_{\pi\pi\mathrm{high}}\bigg)\omega_-\bigg]\bigg\},\\
\end{split}
\end{equation}
where $s_{\pi\pi\mathrm{low(high)}}$ is the low (high) invariant mass squared of mesons $\pi^+$ and $\pi^-$,
\subsection{Resonance contributions}
For the $B^-\rightarrow \pi^-\pi^+\pi^-$ decay, the LHCb Collaboration has studied the $CP$ asymmetries in localized regions of phase space $m_{\pi^+\pi^- \mathrm{high}}^2>15$ $\mathrm{GeV}^2$ and $m_{\pi^+\pi^-\mathrm{low}}^2<0.4$ $\mathrm{GeV}^2$ \cite{Aaij:2013bla}, which include $\sigma(600)$ and $\rho^0(770)$ resonances. For simplicity, these two mesons will be denoted by $\sigma$ and $\rho$, respectively. The total resonance amplitude induced by $\sigma$ and $\rho$ intermediate resonances can be written as
\begin{equation}\label{AR}
\begin{split}
A_R=A_{\rho}+A_{\sigma},
\end{split}
\end{equation}
where
\begin{equation} \label{Hrho}
\begin{split}
A_{\rho}=\frac{\langle \rho \pi^-|\mathcal{H}_{eff}|B^-\rangle \langle \pi^+\pi^-|\mathcal{H}_{\rho \pi^+\pi^-}|\rho\rangle}{S_{\rho}},
 \end{split}
 \end{equation}
\begin{equation} \label{Hf0}
\begin{split}
A_{\sigma}=\langle \sigma \pi^-|\mathcal{H}_{eff}|B^-\rangle \langle \pi^+\pi^-|\mathcal{H}_{\sigma\pi^+\pi^-}|\sigma\rangle R_{\sigma}(s). \\
 \end{split}
 \end{equation}
In Eqs. (\ref{Hrho}, \ref{Hf0}), $\mathcal{H}_{\rho\pi^+\pi^-}=-ig_{\rho\pi\pi}\rho_\mu\pi^+ \overleftrightarrow {\partial}^\mu \pi^-$,
$\mathcal{H}_{\sigma\pi^+\pi^-}=g_{\sigma\pi\pi}\sigma(2\pi^+\pi^-+\pi^0\pi^0)$, where $\rho_\mu$, $\sigma$ and $\pi^\pm$ are the field operators for $\rho^0(770)$, $\sigma(600)$ and $\pi$ mesons, respectively, $g_{\rho\pi\pi}$ and $g_{\sigma\pi\pi}$ are the effective coupling constants which can be expressed in terms of the decay widths of $\rho\rightarrow\pi^+\pi^-$ and $\sigma\rightarrow\pi^+\pi^-$, respectively. In Eq. (\ref{Hrho}), $S_{\rho}$ is the reciprocal of the propagator of $\rho$ and takes the form $s_{\pi\pi\mathrm{low}}-m_{\rho}^2+im_{\rho}\Gamma_{\rho}$, $m_{\rho}$ and $\Gamma_{\rho}$ are the mass and width of the $\rho$ meson, respectively. As for the scalar intermediate state $\sigma$ in Eq. (\ref{Hf0}), we shall adopt the Breit-Wigner function and Bugg model to deal with its propagator, respectively.

With the Breit-Wigner form, $R_{\sigma}(s)$ has the following expression:
\begin{equation} \label{RSBW}
R^{\mathrm{BW}}_{\sigma}(s)=\frac{1}{s_{\pi\pi\mathrm{low}}-m_{\sigma}^2+im_{\sigma}\Gamma_\sigma (s_{\pi\pi\mathrm{low}})},
 \end{equation}
where $\Gamma_\sigma (s_{\pi\pi\mathrm{low}})$ is the width which is a function of $s_{\pi\pi\mathrm{low}}$ and has the expression
\begin{equation}
\begin{split}
\Gamma_\sigma (s_{\pi\pi\mathrm{low}})=\Gamma_\sigma\frac{m_\sigma}{\sqrt{s_{\pi\pi\mathrm{low}}}}\frac{p'(s_{\pi\pi\mathrm{low}})}{p'(m_\sigma^2)},
\end{split}
\end{equation}
where $p'(s_{\pi\pi\mathrm{low}})$ and $p'(m_\sigma^2)$ are the magnitude of the c.m. momenta of $\pi^+$ or $\pi^-$ in the $\pi^+\pi^-$ and $\sigma$ rest frames, respectively, and
\begin{equation}
\begin{split}
p'(s_{\pi\pi\mathrm{low}})&=\frac{1}{2\sqrt{s_{\pi\pi\mathrm{low}}}}\sqrt{s_{\pi\pi\mathrm{low}}(s_{\pi\pi\mathrm{low}}-4m_\pi^2)},\\
p'(m_\sigma^2)&=\frac{1}{2m_\sigma}\sqrt{m_\sigma^2(m_\sigma^2-4m_\pi^2)}.\\
\end{split}
\end{equation}

In the Bugg model, the propagator of $\sigma$ is given by \cite{Bugg:2006gc,Aaij:2015sqa,Li:2015tja}
\begin{equation}\label{T1}
\begin{split}
 R^{\mathrm{Bugg}}_{\sigma}(s)=1/\bigg[M^2-s-g_1^2(s)\frac{s-s_A}{M^2-s_A}z(s)-iM\Gamma_{\mathrm{tot}}(s)\bigg],
 \end{split}
 \end{equation}
 where $z(s)=j_1(s)-j_1(M^2)$ with $j_1(s)=\frac{1}{\pi}[2+\rho_1\ln(\frac{1-\rho_1}{1+\rho_1})]$, $\Gamma_{\mathrm{tot}}(s)=\sum\limits_{i=1}^4 \Gamma_i(s)$ and
 \begin{equation}\label{T2}
\begin{split}
M\Gamma_1(s)&=g_1^2(s)\frac{s-s_A}{M^2-s_A}\rho_1(s),\\
M\Gamma_2(s)&=0.6g_1^2(s)(s/M^2)\mathrm{exp}(-\alpha|s-4m_K^2|)\rho_2(s),\\
M\Gamma_3(s)&=0.2g_1^2(s)(s/M^2)\mathrm{exp}(-\alpha|s-4m_\eta^2|)\rho_3(s),\\
M\Gamma_4(s)&=Mg_{4\pi}\rho_{4\pi}(s)/\rho_{4\pi}(M^2),\\
g_1^2(s)&=M(c_1+c_2s)\mathrm{exp}[-(s-M^2)/A],\\
\rho_{4\pi}(s)&=1.0/[1+\mathrm{exp}(7.082-2.845s)].\\
 \end{split}
 \end{equation}
The parameters in Eqs. (\ref{T1}, \ref{T2}) are fixed to be $M=0.953 \mathrm{GeV}$, $s_A=0.14m_\pi^2$, $c_1=1.302\mathrm{GeV}$,
$c_2=0.340\mathrm{GeV}^{-1}$, $A=2.426\mathrm{GeV}^2$ and $g_{4\pi}=0.011\mathrm{GeV}$ which are given in the fourth column of Table I in Ref. \cite{Bugg:2006gc}. The parameters $\rho_{1,2,3}$ are the phase-space factors of the decay channels $\pi\pi$, $KK$ and $\eta\eta$, respectively, which are defined as \cite{Bugg:2006gc}
\begin{equation}\label{rho}
\rho_i(s)=\sqrt{1-4\frac{m_i^2}{s}},
\end{equation}
with $m_1=m_\pi$, $m_2=m_K$, $m_3=m_\eta$ and $s=s_{\pi\pi\mathrm{low}}$.

Within the QCDF, we can derive the amplitude contributed from $\rho$ resonance to the $B^-\rightarrow \pi^-\pi^+\pi^-$ decay and obtain
\begin{equation}\label{Arho}
\begin{split}
A_\rho&=\frac{-iG_Fg_{\rho\pi\pi}}{S_\rho}(\hat{s}_{\pi\pi\mathrm{high}}-s_{\pi\pi\mathrm{high}})\sum_{p=u,c}\lambda_p^{d}m_\rho\bigg\{\bigg[\delta_{pu}\alpha_1^p(\rho \pi)+\alpha_4^p(\rho \pi)+\alpha_{4,EW}^p(\rho \pi)\bigg]A_0^{B\rightarrow \rho}(0)f_{\pi}\\
&+\bigg[\delta_{pu}\alpha_2^p(\pi \rho)-\alpha_4^p(\pi \rho)+\frac{3}{2}\alpha_{3,EW}^p(\pi \rho)
+\frac{1}{2}\alpha_{3,EW}^p(\pi \rho)\bigg]F_0^{B\rightarrow \pi}(0)f_\rho
+\bigg[\delta_{pu}b_2(\rho \pi)+b_3^p(\rho \pi)\\&
+b_{3,EW}^p(\rho \pi)\bigg]f_Bf_\rho f_{\pi}/(m_Bp_c)+\bigg[-\delta_{pu}b_2(\pi \rho)+b_3^p(\pi \rho)-\frac{1}{2}b_{3,EW}^p(\pi \rho)\bigg]f_Bf_\rho f_{\pi}/(m_Bp_c)\bigg\},\\
  \end{split}
  \end{equation}
for the $B^-\rightarrow\rho \pi^-\rightarrow\pi^+\pi^- \pi^-$ decay mode, where $\hat{s}_{\pi\pi\mathrm{high}}$ is the midpoint of the allowed range of $s_{\pi\pi\mathrm{high}}$, i.e. $\hat{s}_{\pi\pi\mathrm{high}}=(s_{\pi\pi\mathrm{high}, \textrm{max}}+s_{\pi\pi\mathrm{high},\textrm{min}})/2$, with $s_{\pi\pi\mathrm{high},\textrm{max}}$ and $s_{\pi\pi\mathrm{high},\textrm{min}}$ being the maximum and minimum values of $s_{\pi\pi\mathrm{high}}$ for fixed $s_{\pi\pi\mathrm{low}}$.

The amplitude of the  $B^-\rightarrow \sigma\pi^-$ decay can be expressed as
\begin{equation}\label{Asigma0}
\begin{split}
A_\sigma(B\rightarrow \sigma\pi) &=iG_F\sum_{p=u,c}\lambda_p^{d}\bigg\{(m_{\sigma}^2-m_B^2)F_0^{B\rightarrow f}(m_\pi^2)
 f_\pi\bigg[\delta_{pu}\alpha_1(\sigma \pi)+\alpha_4^p(\sigma \pi)+\alpha_{4,EW}^p(\sigma \pi)\bigg]\\
&-f_Bf_\pi\bar{f}^u_{\sigma}\bigg[\delta_{pu}b_2(\sigma \pi)+b_3^p(\sigma \pi)+b_{3,EW}^p(\sigma \pi)\bigg]
 +\bigg[\delta_{pu}\alpha_2(\pi \sigma)+2\alpha_3^p(\pi\sigma)+\alpha_4^p(\pi\sigma)\\
 &+\frac{1}{2}\alpha_{3,EW}^p(\pi\sigma)-\frac{1}{2}\alpha_{4,EW}^p(\pi\sigma)\bigg](m_B^2-m_\pi^2)F_0^{B\rightarrow \pi}(0)\bar{f}^u_{\sigma}
 +\bigg[\sqrt{2}\alpha_3^p(\pi \sigma)-\sqrt{2}\alpha_{3,EW}^p(\pi \sigma)\bigg]\\
 &\times(m_B^2-m_\pi^2)F_0^{B\rightarrow \pi}(0)\bar{f}^s_{\sigma}
 +f_Bf_\pi\bar{f}^u_{\sigma}\bigg[\delta_{pu}b_2(\pi\sigma)+b_3^p(\pi\sigma)-\frac{1}{2}b_{3,EW}^p(\pi\sigma)\bigg]\bigg\}.\\
\end{split}
\end{equation}
If we take the Breit-Wigner (Bugg) model, the amplitude contributed from $\sigma$ resonance to $B^-\rightarrow \pi^-\pi^+\pi^-$ decay can be expressed as
\begin{equation}\label{Asigma2}
\begin{split}
A^{\mathrm{BW}(\mathrm{Bugg})}_\sigma &=2g_{\sigma\pi\pi}R^{\mathrm{BW}(\mathrm{Bugg})}_{\sigma}(s)A_\sigma(B\rightarrow \sigma\pi),\\
\end{split}
\end{equation}

 Then the decay amplitudes of $B^-\rightarrow \rho+\sigma^{\mathrm{BW}(\mathrm{Bugg})}\rightarrow \pi^-\pi^+\pi^-$ can be finally obtained as
\begin{equation}\label{AA}
\begin{split}
A^{\mathrm{BW}(\mathrm{Bugg})}_R=A^{\mathrm{BW}(\mathrm{Bugg})}_\sigma+A_\rho,\\
\end{split}
\end{equation}

\subsection{Total results for the amplitudes of $B^-\rightarrow \pi^-\pi^+\pi^-$ in two different models}
Totally, the decay amplitude for $B\rightarrow \pi^-\pi^+\pi^-$ is the sum of resonance ($R$) contributions and the nonresonance ($NR$) background \cite{Cheng:2013dua}
\begin{equation}\label{A}
A=\sum_R A_R+A_{NR}.
 \end{equation}

In the QCDF, both the resonance and nonresonance contributions have been considered. The decay amplitude via $B^-\rightarrow R^{\mathrm{BW}(\mathrm{Bugg})}+NR\rightarrow \pi^-\pi^+\pi^-$ can be written as

\begin{equation}\label{Atotal11}
\begin{split}
A^{\mathrm{BW}(\mathrm{Bugg})}&=iG_Fg_{\sigma\pi\pi}R^{\mathrm{BW}(\mathrm{Bugg})}_{\sigma}(s)\sum_{p=u,c}\lambda_p^{d}\bigg\{(m_{\sigma}^2-m_B^2)F_0^{B\rightarrow f}(m_\pi^2)
 f_\pi\bigg[\delta_{pu}\alpha_1(\sigma \pi)+\alpha_4^p(\sigma \pi)+\alpha_{4,EW}^p(\sigma \pi)\bigg]\\
 &-f_Bf_\pi\bar{f}^u_{\sigma}\bigg[\delta_{pu}b_2(\sigma \pi)+b_3^p(\sigma \pi)+b_{3,EW}^p(\sigma \pi)\bigg]
 +\bigg[\delta_{pu}\alpha_2(\pi \sigma)+2\alpha_3^p(\pi\sigma)+\alpha_4^p(\pi\sigma)+\frac{1}{2}\alpha_{3,EW}^p(\pi\sigma)\\
 &-\frac{1}{2}\alpha_{4,EW}^p(\pi\sigma)\bigg](m_B^2-m_\pi^2)F_0^{B\rightarrow \pi}(0)\bar{f}^u_{\sigma}
 +\bigg[\sqrt{2}\alpha_3^p(\pi \sigma)-\sqrt{2}\alpha_{3,EW}^p(\pi \sigma)\bigg](m_B^2-m_\pi^2)F_0^{B\rightarrow \pi}(0)\bar{f}^s_{\sigma}\\
 &+f_Bf_\pi\bar{f}^u_{\sigma}\bigg[\delta_{pu}b_2(\pi\sigma)+b_3^p(\pi\sigma)-\frac{1}{2}b_{3,EW}^p(\pi\sigma)\bigg]\bigg\}+\frac{-iG_Fg_{\rho\pi\pi}}{S_\rho}(\hat{s}_{\pi\pi\mathrm{high}}-s_{\pi\pi\mathrm{high}})\sum_{p=u,c}\lambda_p^{d}m_\rho\\
 &\times\bigg\{\bigg[\delta_{pu}\alpha_1^p(\rho \pi)+\alpha_4^p(\rho \pi)+\alpha_{4,EW}^p(\rho \pi)\bigg]A_0^{B\rightarrow \rho}(0)f_{\pi}+\bigg[\delta_{pu}\alpha_2^p(\pi \rho)-\alpha_4^p(\pi \rho)+\frac{3}{2}\alpha_{3,EW}^p(\pi \rho)\\
 &+\frac{1}{2}\alpha_{3,EW}^p(\pi \rho)\bigg] F_0^{B\rightarrow \pi}(0)f_\rho+\bigg[\delta_{pu}b_2(\rho \pi)+b_3^p(\rho \pi)+b_{3,EW}^p(\rho \pi)\bigg]f_Bf_\rho f_{\pi}/(m_Bp_c)+\bigg[-\delta_{pu}b_2(\pi \rho)\\
 &+b_3^p(\pi \rho)-\frac{1}{2}b_{3,EW}^p(\pi \rho)\bigg]f_Bf_\rho f_{\pi}/(m_Bp_c)\bigg\}+\frac{G_F}{\sqrt{2}}\sum_{p=u,c}\lambda_{p}^{d}\bigg\{-\frac{f_\pi}{2}\bigg[2m_\pi^2r+(m_B^2-s_{\pi\pi\mathrm{low}}-m_\pi^2)\omega_+\\
 &+\bigg(m_B^2+3m_\pi^2-s_{\pi\pi\mathrm{low}}-2s_{\pi\pi\mathrm{high}}\bigg)\omega_-\bigg]\bigg[a_1\delta_{pu}+a_4^p+a_{10}^p-(a_6^p+a_8^p)r_\chi^\pi\bigg]
e^{-\alpha_\mathrm{NR}(m_B^2-m_\pi^2-s_{\pi\pi\mathrm{low}})/2}\bigg\}.\\
\end{split}
\end{equation}

\subsection{Localizd CP violation of $B^-\rightarrow \pi^-\pi^+\pi^-$ and branching ratio of $B^-\rightarrow \sigma\pi^-$}

The differential CP asymmetry parameter can be defined as
 \begin{equation}\label{CP asymmetry parameter}
\mathcal{A_{CP}}=\frac{|A|^2-|\bar{A}|^2}{|A|^2+|\bar{A}|^2}.
 \end{equation}

In this work, we will consider resonances and nonresonance in a certain region $\Omega$ which includes $m_{\pi^+\pi^- \mathrm{high}}^2>15$ $\mathrm{GeV}^2$ and $m_{\pi^+\pi^-\mathrm{low}}^2<0.4$ $\mathrm{GeV}^2$ for the $B^{-}\rightarrow \pi^-\pi^+\pi^-$ decay. By integrating the denominator and numerator of $\mathcal{A_{CP}}$ in this region, we get the localized integrated $CP$ asymmetry, which can be measured by experiments and takes the following form:
  \begin{equation}\label{localized CP}
\mathcal{A^\mathrm{\Omega}_{CP}}=\frac{\int_\Omega ds_{12}ds_{13}(|A|^2-|\bar{A}|^2)}{\int_\Omega ds_{12}ds_{13}(|A|^2+|\bar{A}|^2)}.
 \end{equation}

The branching fraction of the $B\rightarrow M_1M_2$ decay has the following form:
\begin{equation}\label{e1}
\mathcal{B}(B\rightarrow M_1M_2)=\tau_B \frac{p_c}{8\pi m_B^2}|A(B\rightarrow M_1M_2)|^2,
\end{equation}
where $\tau_B$ and $m_B$ are the lifetime and the mass of $B$ meson, respectively, $p_c$ is the magnitude of the three momentum of either final state meson in the rest frame of the $B$ meson which can be expressed as
\begin{equation}\label{e2}
p_c=\frac{1}{2m_B}\sqrt{[m_B^2-(m_1+m_2)^2][m_B^2-(m_1-m_2)^2]},
\end{equation}
with $m_1$ and $m_2$ being the two final state mesons' masses, respectively.

The decay rate of the resonance three-body decay is given by \cite{Cheng:2002mk}
\begin{equation}\label{GammaB1}
\begin{split}
\Gamma(B\rightarrow SM_3\rightarrow M_1M_2M_3)&=\frac{1}{2m_B}\int_{(m_1+m_2)^2}^{(m_B-m_3)^2}\frac{ds}{2\pi}|M(B\rightarrow SM_3)|^2|R_S|^2 g_{SM_1M_2}^2\\
&\times\frac{\lambda^{1/2}(m_B^2,s,m_3^2)}{8\pi m_B^2}\frac{\lambda^{1/2}(s,m_1^2,m_3^2)}{8\pi s},\\
\end{split}
\end{equation}
via a scalar resonance, where $s=s_{\pi\pi\mathrm{low}}$, $\lambda$ is the usual triangular function $\lambda(a,b,c)=a^2+b^2+c^2-2ab-2bc-2ca$, and $g_{SM_1M_2}$ is the strong coupling constant which can can be determine from the measured width of the scalar resonance. 

When we use the Breit-Wigner form to deal with the scalar meason, the decay rate in Eq. (\ref{GammaB1}) becomes
\begin{equation}\label{GammaB2}
\begin{split}
\Gamma(B\rightarrow SM_3\rightarrow M_1M_2M_3)&=\frac{1}{2m_B}\int_{(m_1+m_2)^2}^{(m_B-m_3)^2}\frac{ds}{2\pi}|M(B\rightarrow SM_3)|^2 \frac{\lambda^{1/2}(m_B^2,s,m_3^2)}{8\pi m_B^2}\\
&\times\frac{1}{(s-m_S^2)^2+m_S^2\Gamma_S^2(s)}g_{S \pi\pi}^2\frac{\lambda^{1/2}(s,m_1^2,m_3^2)}{8\pi s}.\\
\end{split}
\end{equation}

If the resonance width $\Gamma_S$ is narrow, the expression of the resonance decay rate can be simplified by applying the so-called narrow width approximation
\begin{equation}\label{nwa}
\frac{1}{(s-m_S^2)^2+m_S^2\Gamma_S^2(s)}\approx \frac{\pi}{m_S\Gamma_S}\delta(s-m_S^2).
\end{equation}
Noting
\begin{equation}\label{gammaBSM}
\begin{split}
\Gamma(B\rightarrow SM_3)&=|\langle SM_3|\mathcal{H}_{eff}|B\rangle|^2\frac{p}{8\pi m_B^2},\\
\Gamma(S\rightarrow M_1M_2)&=g_{SM_1M_2}^2\frac{p'(m_S^2)}{8\pi m_S^2},\\
\end{split}
\end{equation}
where $p=\lambda^{1/2}(m_B^2,m_S^2,m_P^2)/(2m_B)$ is the magnitude of the c.m. three momentum of final state particles in the $B$ rest frame and $p'(m_S^2)=\lambda^{1/2}(m_S^2,m_1^2,m_2^2)/(2m_S)$, Eq. (\ref{nwa}) leads to the following ``factorization" relation \cite{Cheng:2002mk}:
\begin{equation}\label{gammaBSM3}
\Gamma(B\rightarrow SM_3\rightarrow M_1M_2M_3)=\Gamma(B\rightarrow SM_3)\mathcal{B}(S\rightarrow M_1M_2),
\end{equation}
and hence,
\begin{equation}\label{mathcalBBSM3}
\mathcal{B}(B\rightarrow SM_3\rightarrow M_1M_2M_3)=\mathcal{B}(B\rightarrow SM_3)\mathcal{B}(S\rightarrow M_1M_2).
\end{equation}

In fact, this factorization relation works reasonably well as long as the two body decay $B\rightarrow SP$ is kinematically allowed and the resonance is narrow. However,  when $B\rightarrow S M_3$ happens at the margin of the  kinematically allowed region or is even not allowed, the off resonance peak effect of the intermediate resonance state will become important and we should consider the finite width effect of $S$ meson. Since the width of the $\sigma$ resonance is very broad, there is no reason to neglect its finite width effect.

We shall follow Refs. \cite{Cheng:2002mk,Cheng:2010vk} to define the parameter $\eta$:
\begin{equation}\label{yita1}
\eta=\frac{\Gamma(B\rightarrow SP\rightarrow M_1M_2M_3)}{\Gamma(B\rightarrow SP)\mathcal{B}(S\rightarrow M_1M_2)}.
\end{equation}

Then we can easily derive the branching ratio of $B\rightarrow SP$:
\begin{equation}\label{BBSP}
\mathcal{B}(B\rightarrow SP)=\frac{\mathcal{B}(B\rightarrow SP\rightarrow M_1M_2M_3)}{\eta\mathcal{B}(S\rightarrow M_1M_2)}.
\end{equation}

The deviation of $\eta$ from unity will give a measure of the violation of the factorization relation in Eq. (\ref{gammaBSM3}). $\eta$ has the following expression:
\begin{equation}\label{yita3}
\begin{split}
\eta^{\mathrm{BW}}=\frac{m_S^2}{4\pi m_B}\frac{\Gamma_S}{pp'(m_S^2)}\int_{(m_1+m_2)^2}^{(m_B-m_3)^2}\frac{ds}{s}\lambda^{1/2}(m_B^2,s,m_3^2)\lambda^{1/2}(s,m_1^2,m_2^2)\frac{1}{(s-m_S^2)^2+(\Gamma_{12}(s)m_S)^2}.
\end{split}
\end{equation}

Similarly, if we adopt the propagator of the $\sigma$ resonance in the Bugg model, the decay rate of the resonance three-body decay can be expressed as
\begin{equation}\label{yita2}
\begin{split}
\Gamma(B\rightarrow SM_3\rightarrow M_1M_2M_3)&=\frac{1}{2m_B}\int_{(m_1+m_2)^2}^{(m_B-m_3)^2}\frac{ds}{2\pi}|M(B\rightarrow SM_3)|^2\frac{\lambda^{1/2}(m_B^2,s,m_3^2)}{8\pi m_B^2}\\
&\times\frac{1}{(M^2-s-g_1^2\frac{s-s_A}{M^2-s_A}z(s))^2+M^2\Gamma^2_{\mathrm{tot}}(s)}g_{SM_1M_2}^2\frac{\lambda^{1/2}(s,m_1^2,m_3^2)}{8\pi s},\\
\end{split}
\end{equation}
combining Eqs. (\ref{gammaBSM}), (\ref{yita2}) and Eq. (\ref{yita1}), we can get
\begin{equation}\label{yita4}
\begin{split}
\eta^{\mathrm{Bugg}}&=\frac{m_S^2}{4\pi m_B}\frac{\Gamma_S}{pp'(m_S^2)}\int_{(m_1+m_2)^2}^{(m_B-m_3)^2}\frac{ds}{s}\lambda^{1/2}(m_B^2,s,m_3^2)\lambda^{1/2}(s,m_1^2,m_2^2)\\
&\times\frac{1}{(M^2-s-g_1^2\frac{s-s_A}{M^2-s_A}z(s))^2+M^2\Gamma^2_{\mathrm{tot}}(s)}.\\
\end{split}
\end{equation}
\section{Numerical results}
The theoretical results obtained in the QCDF approach depend on many input parameters. The values of the Wolfenstein parameters are given as $\bar{\rho}=0.117\pm0.021$, $\bar{\eta}=0.353\pm0.013$ \cite{Agashe:2014kda}.

The effective Wilson coefficients used in our calculations are taken from Ref. \cite{Wang:2014hba}:
\begin{equation}\label{C}
\begin{split}
&c'_1=-0.3125, \quad c'_2=1.1502, \\
&c'_3=2.433\times10^{-2}+1.543\times10^{-3}i,\quad c'_4=-5.808\times10^{-2}-4.628\times10^{-3}i, \\
&c'_5=1.733\times10^{-2}+1.543\times10^{-3}i,\quad c'_6=-6.668\times10^{-2}-4.628\times10^{-3}i, \\
&c'_7=-1.435\times10^{-4}-2.963\times10^{-5}i,\quad c'_8=3.839\times10^{-4}, \\
&c'_9=-1.023\times10^{-2}-2.963\times10^{-5}i,\quad c'_{10}=1.959\times10^{-3}. \\
\end{split}
\end{equation}

For the masses appeared in $B$ decays, we shall use the following values \cite{Agashe:2014kda} (in units of $\mathrm{GeV}$):
\begin{equation}
\begin{split}
m_u&=m_d=0.0035,\quad m_s=0.119, \quad m_b=4.2,\quad m_q=\frac{m_u+m_d}{2},\quad m_{\pi^\pm}=0.14,\\
m_{B^-}&=5.279,\quad m_\rho=0.775,\quad m_\sigma=0.5,\\
\end{split}
\end{equation}
while for the widthes we shall use (in units of $\mathrm{GeV}$) \cite{Agashe:2014kda}
\begin{equation}
\Gamma_{\rho}=0.149,\quad\Gamma_{\rho\rightarrow\pi\pi}=0.149,\quad\Gamma_{\sigma}=0.5,\quad\Gamma_{\sigma\rightarrow\pi\pi}=0.335.\\
\end{equation}

The following numerical values for the decay constants will be used \cite{Cheng:2013dua,Cheng:2010yd,Cheng:2005nb} (in units of $\mathrm{GeV}$):
\begin{equation}
\begin{split}
f_{\pi^\pm}&=0.131,\quad f_{B^-}=0.21\pm0.02, \quad f_{K^-}=0.156\pm0.007, \quad \bar{f}_\sigma^u=0.4829\pm0.14,  \\
 \bar{f}^s_\sigma&=-0.21\pm0.10,\quad f_\rho=0.216\pm0.003,\quad f_\rho^\perp =0.165\pm0.009.\\
\end{split}
\end{equation}

As for the form factors, we use \cite{Cheng:2013dua,Cheng:2010yd,Cheng:2005nb}
\begin{equation}
\begin{split}
F_0^{B\rightarrow K}(0)&=0.35\pm0.04,\quad F_0^{B\rightarrow \sigma}(m_K^2)=0.45\pm0.15,\\
A_0^{B\rightarrow \rho}(0)&=0.303\pm0.029,\quad F_0^{B\rightarrow \pi}(0)=0.25\pm0.03.\\
\end{split}
\end{equation}

The values of Gegenbauer moments at $\mu=1 \mathrm{GeV}$ are taken from \cite{Cheng:2013dua,Cheng:2010yd,Cheng:2005nb}:
\begin{equation}
\begin{split}
\alpha_1^\rho&=0,\quad \alpha_2^\rho=0.15\pm0.07, \quad \alpha_{1,\perp}^\rho=0,\quad \alpha_{2,\perp}^\rho=0.14\pm0.06, \\
B_{1,\sigma}^u&=-0.42\pm0.074,\quad B_{3,\sigma}^u=-0.58\pm0.23,\\
 B_{1,\sigma}^s&=-0.35\pm0.061,\quad B_{3,\sigma}^s=-0.43\pm0.18.\\
\end{split}
\end{equation}

Because of the broad width of the $\sigma$ meson, it is not appropriate to deal with the branching fraction of the $B^-\rightarrow\sigma\pi^-$ decay using Eq. (\ref{mathcalBBSM3}). In comparison, we should consider its finite width effect and introduce the parameter $\eta$ defined in Eq. (\ref{yita1}) to modify the branching fraction relationship between the two-body and three-body decays of the $B$ meson. From Eqs. (\ref{yita3}) and (\ref{yita4}), one can obtain $\eta^{\mathrm{BW}}=3.68$ and $\eta^{\mathrm{Bugg}}=0.316$ using the Breit-Wigner form and the Bugg model for the $\sigma$ meson, respectively. In 2005, $BABAR$ Collaboration reported the upper limit of $\mathcal{B}(B^-\rightarrow \sigma \pi^-,\sigma\rightarrow \pi^+\pi^-)$ as $4.1\times10^{-6}$ \cite{Aubert:2005sk}. Inserting this experimental result and the values of $\eta^{\mathrm{BW}}$ and $\eta^{\mathrm{Bugg}}$ into Eq. (\ref{BBSP}), we can obtain the upper limits of $\mathcal{B}(B^-\rightarrow \sigma \pi^-)$ in two different models. As for the value of the branching fraction of the $\sigma\rightarrow\pi^+\pi^-$ decay, we will take $\mathcal{B}(\sigma\rightarrow \pi^+\pi^-)=\frac{2}{3}$ \cite{Cheng:2002mk}. If we use the the Breit-Wigner result $\eta^{\mathrm{BW}}=3.68$, the upper limit of $\mathcal{B}(B^-\rightarrow \sigma \pi^-)$ is $1.67\times10^{-6}$. Likewise, we can also obtain $\mathcal{B}(B^-\rightarrow \sigma \pi^-)<1.946\times10^{-5}$ if we use the Bugg model result $\eta^{\mathrm{Bugg}}=0.316$. Obviously, the upper limits for $\mathcal{B}(B^-\rightarrow \sigma \pi^-)$ are very different in these two different models, so it is significant to study the $\sigma$ meson distribution effects in the Breit-Wigner and the Bugg models even though the finite width effects are considered in both models.

Besides, in 2013, the LHCb Collaboration observed the large $CP$ asymmetry in the localized region of the phase space \cite{Aaij:2013bla}, $\mathcal{A_{CP}}(B^-\rightarrow\pi^-\pi^+\pi^-)=0.584\pm0.082\pm0.027\pm0.007$, for $m_{\pi^+\pi^- \mathrm{high}}^2>15$ $\mathrm{GeV}^2$ and $m_{\pi^+\pi^-\mathrm{low}}^2<0.4$ $\mathrm{GeV}^2$. Generally, a fit of the divergence parameters $\rho$ and $\phi$ to the $B\rightarrow VP$ and $B\rightarrow PV$ data indicates $X^{PV}\neq X^{VP}$, i.e. $\rho^{PV}=0.87$, $\rho^{VP}=1.07$, $\phi^{VP}=-30^0$ and $\phi^{PV}=-70^0$ \cite{Cheng:2009cn}. We shall assign an error of $\pm0.1$ to $\rho^{M_1M_2}$ and $\pm20^0$ to $\phi^{M_1M_2}$ for estimation of theoretical uncertainties. However, for $B\rightarrow PS$ and $B\rightarrow SP$ decays, there is little experimental data so the values of $\rho_S$ and $\phi_S$ are not determined well, we will adopt $X^{PS}=X^{SP}=(1+\rho_S e^{i\phi_S})\ln\frac{m_B}{\Lambda_h}$ as in our previous work \cite{Qi:2018wkj,Qi:2018syl}.  We can get the expressions of  $\mathcal{A_{CP}}(B^-\rightarrow R^\mathrm{BW}+NR\rightarrow \pi^-\pi^+\pi^-)$ and $\mathcal{A_{CP}}(B^-\rightarrow R^\mathrm{Bugg}+NR\rightarrow \pi^-\pi^+\pi^-)$ which are the functions of $\rho_S$ and $\phi_S$. Meanwhile, one can also get the expressions of $\mathcal{B}(B^-\rightarrow \sigma \pi^-)$ in these two models, which are also the functions of $\rho_S$ and $\phi_S$. By fitting the Breit-Wigner model theoretical results of $\mathcal{A_{CP}}(B^-\rightarrow R^\mathrm{BW}+NR\rightarrow \pi^-\pi^+\pi^-)$ and $\mathcal{B}(B^-\rightarrow \sigma \pi^-)$ to the experimental data $\mathcal{A_{CP}}(B^-\rightarrow\pi^-\pi^+\pi^-)=0.584\pm0.082\pm0.027\pm0.007$ in the aforementioned region and $\mathcal{B}(B^-\rightarrow \sigma \pi^-)<1.67\times10^{-6}$, respectively, and setting $\phi_S\in[0,2\pi]$ and $\rho_S\in[0,8]$ \cite{Bobeth:2014rra,Ciuchini:2002uv}, we find there is no allowed $\rho_S$ and $\phi_S$ to satisfy the above two data simultaneously. However, if we use the Bugg model results to fit the experimental data for $\mathcal{A_{CP}}(\pi^-\pi^+\pi^-)$ and $\mathcal{B}(B^-\rightarrow \sigma \pi^-)<1.946\times10^{-5}$, we can get the region $\rho_S\in[1.70,3.34]$ and $\phi_S \in [0.50,4.50]$ which can satisfy these two requirements simultaneously. This indicates that the Bugg model is more appropriate than the Breit-Wigner model when dealing with the broad scalar meson $\sigma$ even though in both models the finite width effects of $\sigma$ are considered. In the Bugg model, the large localized $CP$ violation for the $B^-\rightarrow \pi^-\pi^+\pi^-$ decay can indeed be explained by the interference of the resonances plus the nonresonance contribution.

It is noted that the range of $\rho_S\in[1.70,3.34]$ is larger than the previously conservative choice of $\rho\leq1$ \cite{Beneke:2001ev,Beneke:2003zv}. Since the QCDF itself cannot give information about the parameters $\rho$ and $\phi$, there is no reason to restrict $\rho$ to the range $\rho\leq1$ \cite{Cheng:2009cn,Chang:2014rla,Sun:2014tfa}. In fact, there are many experimental studies which have been successfully carried out at $B$ factories ($BABAR$ and Belle), Tevatron (CDF and D0) and LHCb in the past and will be continued at LHCb and Belle experiments. These experiments provide highly fertile ground for theoretical studies and have yielded many exciting and important results, such as measurements of pure annihilation $B_s\rightarrow \pi \pi$ and $B_d\rightarrow K K$ decays reported recently by CDF, LHCb and Belle \cite{Aaltonen:2011jv,Aaij:2012as,Duh:2012ie}, which suggest the existence of unexpected large annihilation contributions and have attracted much attention \cite{Xiao:2011tx,Gronau:2012gs,Chang:2014rla}. Besides, there are many theoretical studies indicating possible unnegligible large weak annihilation contributions within different approaches in different decays \cite{Keum:2000wi,Qi:2018wkj,Qi:2018syl, Lu:2000em, Cheng:2009cn,Chang:2014rla,Sun:2014tfa}. Therefore, larger values of the $\rho_S$ are acceptable when dealing with the divergence problems for $B\rightarrow SP(PS)$ decays. Much more experimental and theoretical efforts are expected to understand the underlying QCD dynamics of annihilation and spectator scattering contributions.

\section{SUMMARY}
In this work, we study the localized $CP$ violation in the $B^-\rightarrow \pi^-\pi^+\pi^-$ decay and the branching fraction of the $B^-\rightarrow \sigma(600)\pi^-$ decay within the QCD factorization approach.
Both the resonance and nonresonance contributions are included when we study the localized $CP$ asymmetry for the $B^-\rightarrow \pi^-\pi^+\pi^-$ decay. As for the resonance contributions we consider the scalar meson $\sigma(600)$ and the vector meson $\rho^0(770)$. Meanwhile, we adopt the Breit-Wigner and the Bugg models to parameterize the propagator of the $\sigma(600)$ meson, respectively. Since the width of the $\sigma$ resonance is broad, it is necessary to take into account its finite width effect and we follow Refs. \cite{Cheng:2002mk,Cheng:2010vk} to introduce a parameter $\eta$ defined in Eq. (\ref{yita1}) to modify the branching fraction relationship between the two-body and three-body decays of the $B$ meson. From our calculations, we get $\eta^{\mathrm{BW}}=3.68$ and $\eta^{\mathrm{Bugg}}=0.316$ for the Breit-Wigner form and the Bugg model, respectively. Meanwhile, we obtain the upper limits of $\mathcal{B}(B^-\rightarrow \sigma \pi^-)$ as $1.67\times10^{-6}$ and $1.946\times10^{-5}$ in these two different models, respectively. It is worth noting that these two limits are very different, so it is important to study the $\sigma$ meson distribution effects in the Breit-Wigner and the Bugg models even though the finite width effects are considered in both models. By fitting the theoretical results for $\mathcal{A_{CP}}(B^-\rightarrow R+NR\rightarrow \pi^-\pi^+\pi^-)$ and $\mathcal{B}(B^-\rightarrow \sigma \pi^-)$ in these two models to the experimental data $\mathcal{A_{CP}}(\pi^-\pi^+\pi^-)=0.584\pm0.082\pm0.027\pm0.007$ and the upper limits of $\mathcal{B}(B^-\rightarrow \sigma \pi^-)$, we find that there is no allowed $\rho_S$ and $\phi_S$ to be found in the Breit-Wigner model, but there exists the region $\rho_S\in[1.70,3.34]$ and $\phi_S \in [0.50,4.50]$ in the Bugg model. This indicates that the Bugg model is more plausible than the Breit-Wigner model when dealing with the broad scalar meson $\sigma$, since in the Bugg model, the large localized $CP$ violation for the $B^-\rightarrow \pi^-\pi^+\pi^-$ decay can be explained by the interference of the resonances plus the nonresonance contribution. Furthermore, large values of $\rho_S$, $\rho_S\in[1.70,3.34]$, reveal that the contributions from the weak annihilation and the hard spectator scattering processes are both large for the $B^-\rightarrow \pi^-\pi^+\pi^-$ and the $B^-\rightarrow\sigma\pi^-$ decays.  Especially, the contribution from the weak annihilation processes should not be neglected. In fact, the presence of the large weak annihilation and hard spectator scattering contributions has been supported by recent studies both experimentally and theoretically. So the larger values of $\rho_S$ are acceptable when dealing with the divergence problems for the $B\rightarrow SP(PS)$ decays.

\acknowledgments
 This work was supported by National Natural Science Foundation of China (Projects No. 11575023, No. 11775024, No. 11705081, No.11805012).


\begin{thebibliography}{99}

\bibitem{Christenson:1964fg}
  J.~H.~Christenson, J.~W.~Cronin, V.~L.~Fitch and R.~Turlay,
 Phys.\ Rev.\ Lett.\  {\bf 13}, 138 (1964).

\bibitem{Cabibbo:1963yz}
  N.~Cabibbo,
  Phys.\ Rev.\ Lett.\  {\bf 10}, 531 (1963).

\bibitem{Kobayashi:1973fv}
  M.~Kobayashi and T.~Maskawa,
  Prog.\ Theor.\ Phys.\  {\bf 49}, 652 (1973).

\bibitem{Cheng:2013dua}
  H.~Y.~Cheng and C.~K.~Chua,
  Phys.\ Rev.\ D {\bf 88}, 114014 (2013).

\bibitem{Aaij:2013bla}
  R.~Aaij {\it et al.} [LHCb Collaboration],
  Phys.\ Rev.\ Lett.\  {\bf 112}, no. 1, 011801 (2014).

\bibitem{Aubert:2005sk}
  B.~Aubert {\it et al.} [BaBar Collaboration],
  Phys.\ Rev.\ D {\bf 72}, 052002 (2005).

\bibitem{Wirbel:1985ji}
  M.~Wirbel, B.~Stech and M.~Bauer,
  Z.\ Phys.\ C {\bf 29}, 637 (1985).

\bibitem{Bauer:1986bm}
  M.~Bauer, B.~Stech and M.~Wirbel,
  Z.\ Phys.\ C {\bf 34}, 103 (1987).


\bibitem{Beneke:2003zv}
  M.~Beneke and M.~Neubert,
  Nucl.\ Phys.\ B {\bf 675}, 333 (2003).

\bibitem{Beneke:2001ev}
  M.~Beneke, G.~Buchalla, M.~Neubert and C.~T.~Sachrajda,
  Nucl.\ Phys.\ B {\bf 606}, 245 (2001).

\bibitem{Keum:2000ph}
  Y.~Y.~Keum, H.~n.~Li and A.~I.~Sanda,
  Phys.\ Lett.\ B {\bf 504}, 6 (2001).

\bibitem{Bauer:2000ew}
  C.~W.~Bauer, S.~Fleming and M.~E.~Luke,
  Phys.\ Rev.\ D {\bf 63}, 014006 (2000).

\bibitem{Cheng:2010yd}
  H.~Y.~Cheng and K.~C.~Yang,
  Phys.\ Rev.\ D {\bf 83} (2011) 034001.

\bibitem{Cheng:2005nb}
  H.~Y.~Cheng, C.~K.~Chua and K.~C.~Yang,
  Phys.\ Rev.\ D {\bf 73}, 014017 (2006).


\bibitem{Cheng:2007st}
  H.~Y.~Cheng, C.~K.~Chua and K.~C.~Yang,
  Phys.\ Rev.\ D {\bf 77}, 014034 (2008).

\bibitem{Buchalla:1995vs}
  G.~Buchalla, A.~J.~Buras and M.~E.~Lautenbacher,
  Rev.\ Mod.\ Phys.\  {\bf 68} (1996) 1125.

\bibitem{Cheng:2010hn}
  H.~Y.~Cheng, Y.~Koike and K.~C.~Yang,
  Phys.\ Rev.\ D {\bf 82}, 054019 (2010).

\bibitem{Cheng:2009cn}
  H.~Y.~Cheng and C.~K.~Chua,
  Phys.\ Rev.\ D {\bf 80}, 114008 (2009).

\bibitem{Wang:2016yrm}
  C.~Wang, Z.~Y.~Wang, Z.~H.~Zhang and X.~H.~Guo,
  Phys.\ Rev.\ D {\bf 93}, no. 11, 116008 (2016).

 \bibitem{Qi:2018wkj}
  J.~J.~Qi, Z.~Y.~Wang, Z.~H.~Zhang, J.~Xu and X.~H.~Guo,
  Eur.\ Phys.\ J.\ C {\bf 78}, no. 10, 845 (2018).

\bibitem{Qi:2018syl}
  J.~J.~Qi, Z.~Y.~Wang, X.~H.~Guo, Z.~H.~Zhang and C.~Wang,
  arXiv:1811.02167 [hep-ph].

\bibitem{Cheng:2007si}
  H.~Y.~Cheng, C.~K.~Chua and A.~Soni,
  Phys.\ Rev.\ D {\bf 76} (2007) 094006.

\bibitem{Lee:1992ih}
  C.~L.~Y.~Lee, M.~Lu and M.~B.~Wise,
  Phys.\ Rev.\ D {\bf 46}, 5040 (1992).
  
  \bibitem{Fajfer:1998yc}
  S.~Fajfer, R.~J.~Oakes and T.~N.~Pham,
  Phys.\ Rev.\ D {\bf 60}, 054029 (1999).
  
\bibitem{Ahmed:2001xc}
  S.~Ahmed {\it et al.} [CLEO Collaboration], Phys.\ Rev.\ Lett.\  {\bf 87}, 251801 (2001).
  
\bibitem{Yan:1992gz}
  T.~M.~Yan, H.~Y.~Cheng, C.~Y.~Cheung, G.~L.~Lin, Y.~C.~Lin and H.~L.~Yu,
  Phys.\ Rev.\ D {\bf 46}, 1148 (1992).
  

\bibitem{Cheng:2002qu}
  H.~Y.~Cheng and K.~C.~Yang,
  Phys.\ Rev.\ D {\bf 66}, 054015 (2002).











\bibitem{Cheng:2016ajl}
  H.~Y.~Cheng,
  Nucl.\ Part.\ Phys.\ Proc.\  {\bf 273}, 1290 (2016).

\bibitem{Bugg:2006gc}
  D.~V.~Bugg,
  J.\ Phys.\ G {\bf 34}, 151 (2007).


\bibitem{Aaij:2015sqa}
  R.~Aaij {\it et al.} [LHCb Collaboration],
  Phys.\ Rev.\ D {\bf 92}, no. 3, 032002 (2015).


\bibitem{Li:2015tja}
  Y.~Li, A.~J.~Ma, W.~F.~Wang and Z.~J.~Xiao,
  Eur.\ Phys.\ J.\ C {\bf 76}, no. 12, 675 (2016).


\bibitem{Cheng:2002mk}
  H.~Y.~Cheng,
  Phys.\ Rev.\ D {\bf 67}, 054021 (2003).

\bibitem{Cheng:2010vk}
  H.~Y.~Cheng and C.~W.~Chiang,
  Phys.\ Rev.\ D {\bf 81}, 074031 (2010).

\bibitem{Agashe:2014kda}
  K.~A.~Olive {\it et al.} [Particle Data Group],
  Chin.\ Phys.\ C {\bf 38}, 090001 (2014).

\bibitem{Wang:2014hba}
  C.~Wang, X.~H.~Guo, Y.~Liu and R.~C.~Li,
  Eur.\ Phys.\ J.\ C {\bf 74}, no. 11, 3140 (2014).


\bibitem{Bobeth:2014rra}
  C.~Bobeth, M.~Gorbahn and S.~Vickers,
  Eur.\ Phys.\ J.\ C {\bf 75}, no. 7, 340 (2015).


\bibitem{Ciuchini:2002uv}
  M.~Ciuchini, E.~Franco, A.~Masiero and L.~Silvestrini,
  Phys.\ Rev.\ D {\bf 67}, 075016 (2003).

\bibitem{Chang:2014rla}
  Q.~Chang, J.~Sun, Y.~Yang and X.~Li,
  Phys.\ Rev.\ D {\bf 90}, no. 5, 054019 (2014).

\bibitem{Sun:2014tfa}
  J.~Sun, Q.~Chang, X.~Hu and Y.~Yang,
  Phys.\ Lett.\ B {\bf 743}, 444 (2015).

\bibitem{Aaltonen:2011jv}
  T.~Aaltonen {\it et al.} [CDF Collaboration],
  Phys.\ Rev.\ Lett.\  {\bf 108}, 211803 (2012).

\bibitem{Aaij:2012as}
  R.~Aaij {\it et al.} [LHCb Collaboration],
  JHEP {\bf 1210}, 037 (2012).


\bibitem{Duh:2012ie}
  Y.-T.~Duh {\it et al.} [Belle Collaboration],
  Phys.\ Rev.\ D {\bf 87}, no. 3, 031103 (2013).

\bibitem{Xiao:2011tx}
  Z.~J.~Xiao, W.~F.~Wang and Y.~y.~Fan,
  Phys.\ Rev.\ D {\bf 85}, 094003 (2012).

\bibitem{Gronau:2012gs}
  M.~Gronau, D.~London and J.~L.~Rosner,
  Phys.\ Rev.\ D {\bf 87}, no. 3, 036008 (2013).

\bibitem{Keum:2000wi}
  Y.~Y.~Keum, H.~N.~Li and A.~I.~Sanda,
  Phys.\ Rev.\ D {\bf 63}, 054008 (2001).

\bibitem{Lu:2000em}
  C.~D.~Lu, K.~Ukai and M.~Z.~Yang,
  Phys.\ Rev.\ D {\bf 63}, 074009 (2001).

\end{thebibliography}
\end{document}